\documentclass[3p,times,procedia]{elsarticle}
\usepackage{nupha_ecrc}

\volume{00}

\firstpage{1}

\journalname{Nuclear Physics A}

\runauth{A. Aduszkiewicz for the \NASixtyOne Collaboration}

\jid{nupha}

\jnltitlelogo{Nuclear Physics A}

\usepackage{amssymb}

\usepackage{lineno}

\usepackage[figuresright]{rotating}

\usepackage{xspace} 
\xspaceaddexceptions{[]+} 

\usepackage{amsmath}

\newcommand{\avg}[1]{\langle{#1}\rangle}
\newcommand{\dd}{\textup{d}}

\newcommand{\NASixtyOne}{NA61\slash SHINE\xspace}

\newcommand{\dedx}{\ensuremath{\dd E/\dd x}\xspace}
\newcommand{\pt}{\ensuremath{p_\textup{T}}\xspace}
\newcommand{\mt}{\ensuremath{m_\textup{T}}\xspace}

\newcommand{\hm}{\texorpdfstring{\MakeLowercase{\ensuremath{\textup{h}^-}}}{h-}\xspace}
\newcommand{\pip}{\ensuremath{\pi^+}\xspace}
\newcommand{\pim}{\texorpdfstring{\ensuremath{\pi^-}}{pi-}\xspace}

\newcommand{\kp}{\ensuremath{\textup{K}^+}\xspace}

\newcommand{\pbar}{\ensuremath{\bar{\textup{p}}}\xspace}

\newcommand{\eeV}{\text{\texorpdfstring{e\kern-0.03em V}{eV}}}
\newcommand{\overc}{\texorpdfstring{\kern-0.1em /\kern-0.1em \ensuremath{c}}{/c}}
\newcommand{\overcc}{\texorpdfstring{\kern-0.1em /\kern-0.1em \ensuremath{c^2}}{/c}}

\newcommand{\GeVc}{\text{~G\eeV\overc}\xspace}
\newcommand{\GeVcc}{\text{~G\eeV}\overcc\xspace}

\newcommand{\AGeVc}{\ensuremath{A}\text{~G\eeV\overc}\xspace}

\usepackage{hyperref}

\begin{document}

\begin{frontmatter}



\dochead{XXVIth International Conference on Ultrarelativistic Nucleus-Nucleus Collisions\\ (Quark Matter 2017)}

\title{Recent results from \NASixtyOne}


\author{Antoni Aduszkiewicz for the \NASixtyOne Collaboration}

\address{\textcopyright{} 2017. This manuscript version is made available under the CC-BY-NC-ND 4.0 license http://creativecommons.org/licenses/by-nc-nd/4.0/ }

\begin{abstract}
  The \NASixtyOne fixed-target experiment at the CERN SPS studies the onset of deconfinement and searches for the critical point of strongly interacting matter by measuring hadron production as a function of the collision energy and the colliding system size.

  This contribution summarises recent results on hadron spectra and fluctuations, in particular new results on charged kaon production in $^7$Be+$^9$Be collisions.
  Also an overview of the proposed future program of \NASixtyOne is presented.
\end{abstract}


\end{frontmatter}





\section{Two-dimensional scan program of the \NASixtyOne experiment at CERN SPS}

\begin{figure}
  \centering
  \includegraphics[width=0.49\textwidth]{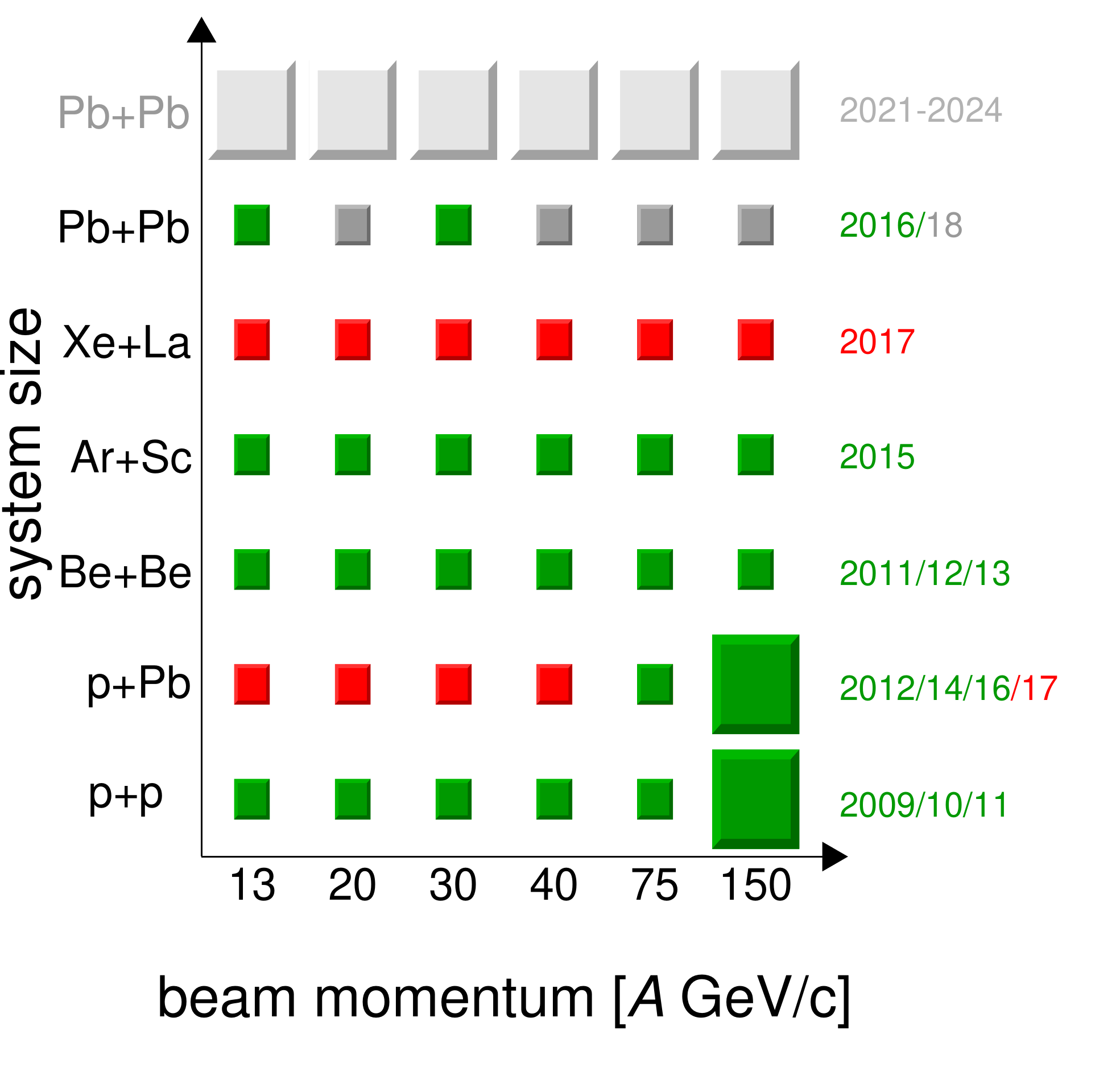}
  \caption{
    Data taking progress of the \NASixtyOne two-dimensional scan.
    The small boxes correspond to $2\cdot 10^6$ events and the large ones to $50\cdot 10^6$.
    The green boxes show data collected as of spring 2017.
    The reactions planned to be measured within the approved and extended \NASixtyOne programs are shown in red and grey, respectively.
    The light grey boxes show the large statistics Pb+Pb beam momentum scan planned for 2021--2024.
  }\label{fig:program}
\end{figure}

\NASixtyOne scans the phase diagram of strongly interacting matter in baryon density and temperature.
The programme is motivated by the evidence for the onset of deconfinement in Pb+Pb collisions at 30\AGeVc found by the NA49 experiment~\cite{Alt:2007aa, Afanasiev:2002mx}.
Measurements of hadron production in a two-dimensional scan in beam momentum (13$A$--150/158\AGeVc) and system size (p+p, p+Pb, $^7$Be+$^9$Be, Ar+Sc, Xe+La and Pb+Pb) are conducted in parallel to the RHIC beam energy scan.
Figure~\ref{fig:program} shows the data taking progress.

\NASixtyOne studies the onset of deconfinement by measurements of the hadron spectra and searches for the critical point of strongly interacting matter by measuring event-by-event fluctuations.

The detector is based on a system of five Time Projection Chambers providing acceptance in the full forward hemisphere, down to $\pt = 0$.
Time of Flight walls provide additional particle identification.
A zero-degree calorimeter, Projectile Spectator Detector, allows the selection of central collisions based on the measurement of the forward energy.

\section{Recent results from \NASixtyOne}
\subsection{Study of the onset of deconfinement}
\subsubsection{Negatively charged pion spectra}
\begin{figure}
  \centering
\includegraphics[width=6.5cm]{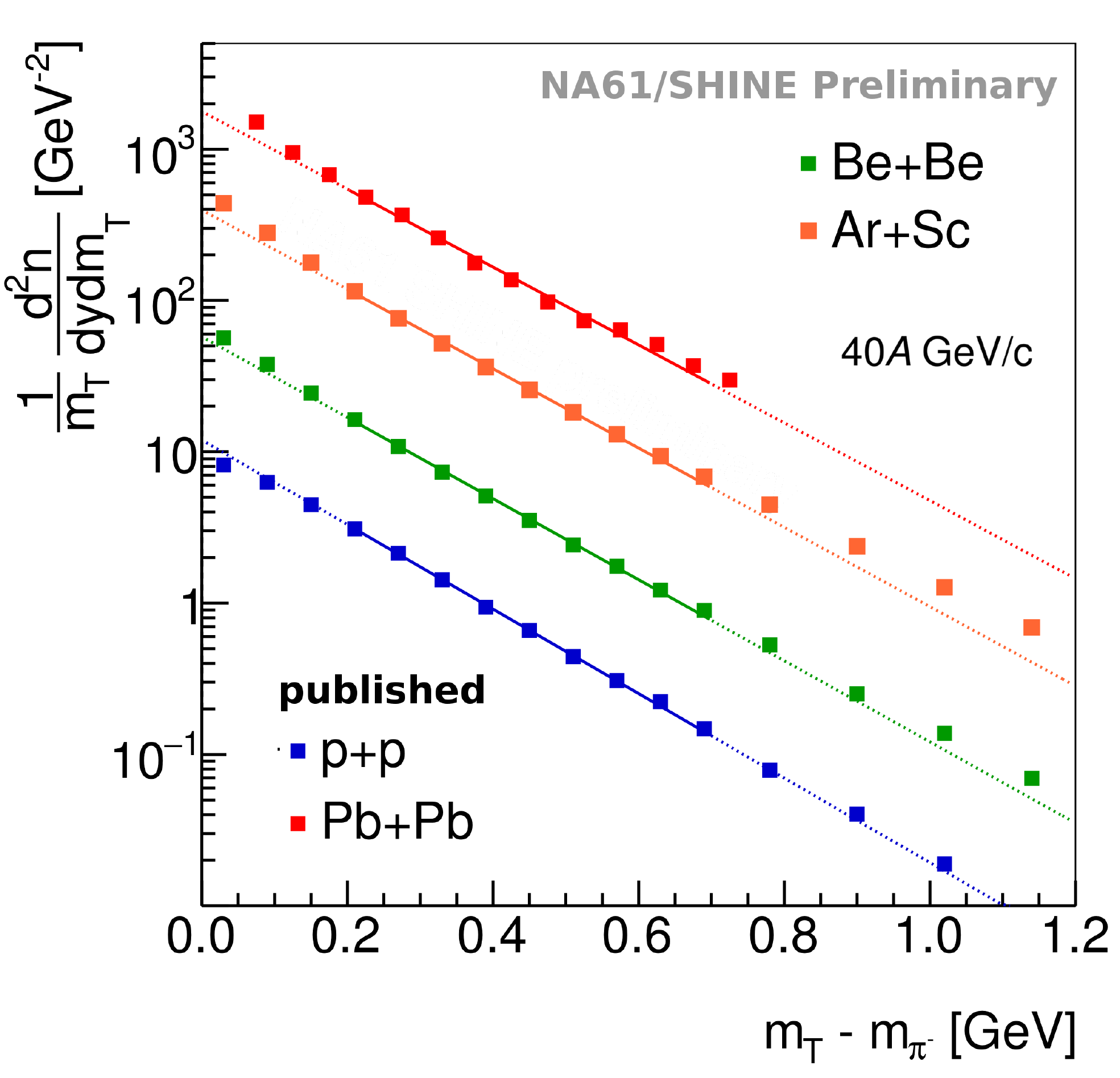}
\includegraphics[width=8.0cm]{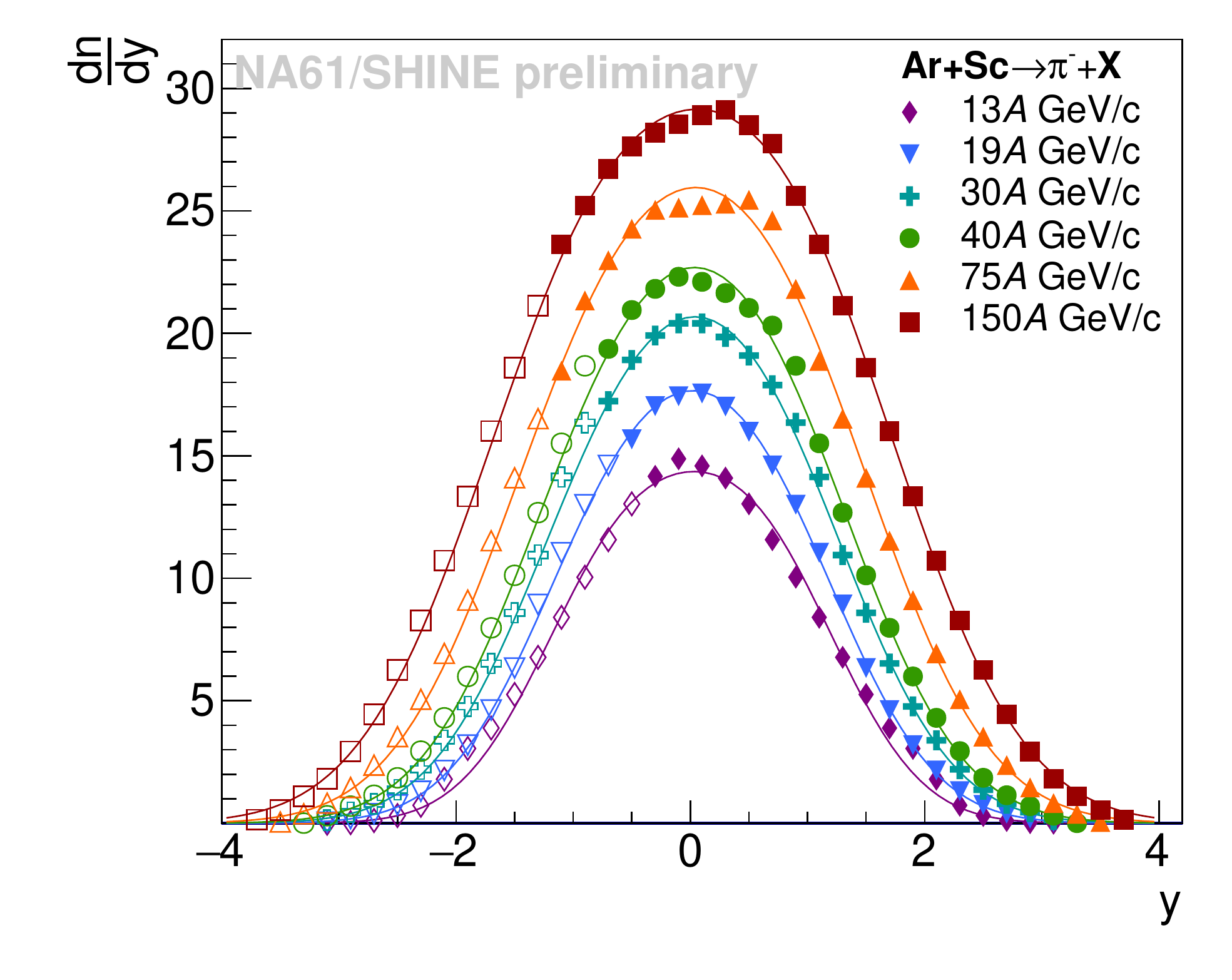}
 \caption{Negatively charged pion spectra.
   \emph{Left:} Transverse mass spectra at mid-rapidity at 40\AGeVc.
   An exponential function was fitted in the region $0.2<\mt<0.7$\GeVcc.
   \emph{Right:} Rapidity spectra in Ar+Sc collisions at six beam momenta.
   A sum of two symmetrically displaced normal distributions of independent amplitudes was fitted to the data.
 }\label{fig:pim_spectra}
\end{figure}

\begin{figure}
  \centering
\includegraphics[width=0.49\textwidth]{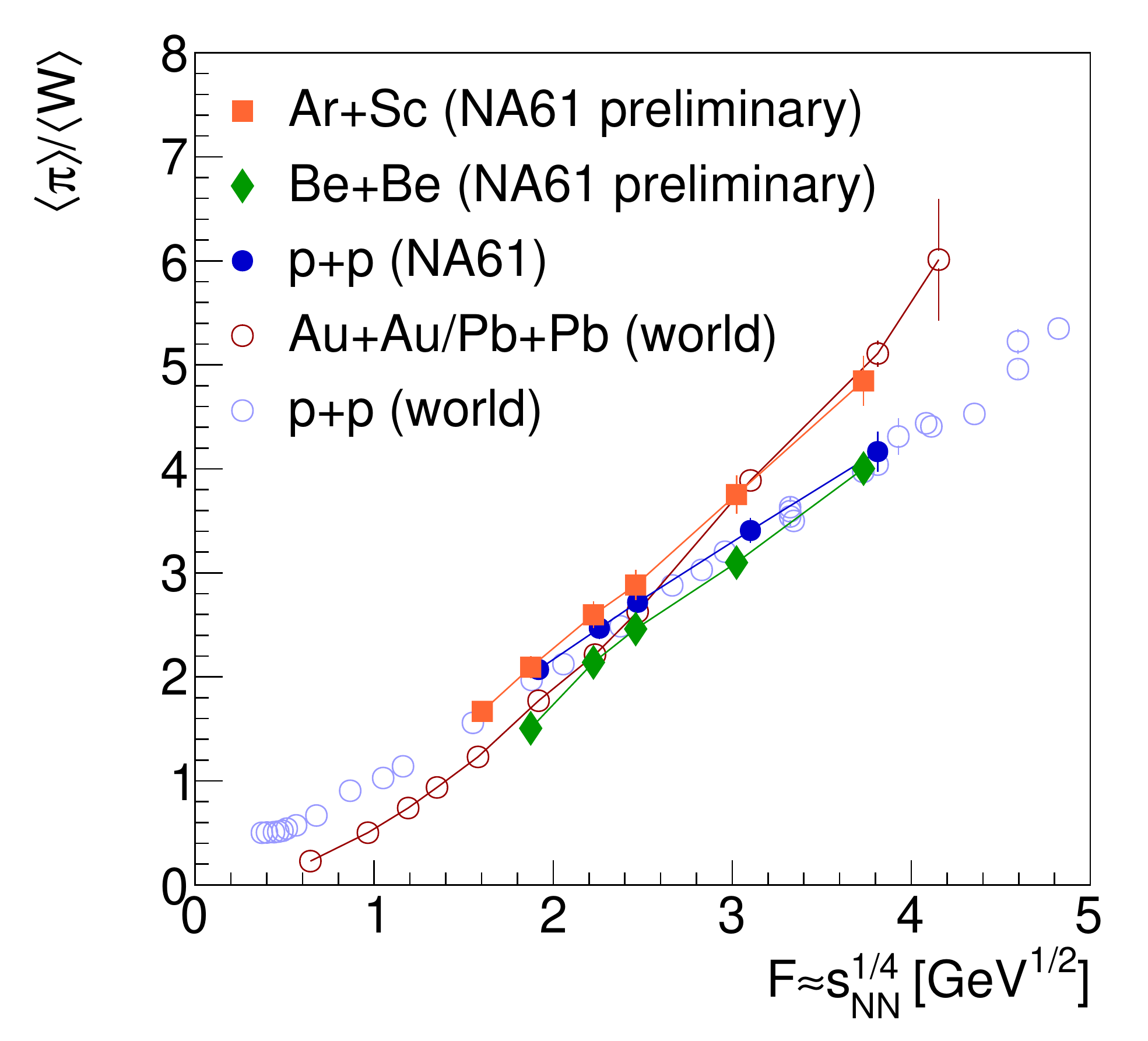}
 \caption{Energy dependence of the total $\pi$ multiplicity in full $4\pi$ phase space calculated based on \pim measurements.
   The data was divided by the number of wounded nucleons $W$.}\label{fig:pim_multiplicity}
\end{figure}

Negatively charged pion spectra in p+p \cite{Abgrall:2013qoa}, central Be+Be \cite{Kaptur:2015xzu, Pulawski:2015rma} and central Ar+Sc collisions \cite{Naskret:2016lut, Lewicki:2016wih, Grebieszkow:2016cza} were derived in large acceptance from unidentified negatively charged hadron spectra using the \hm method.
Figure~\ref{fig:pim_spectra} (\emph{left}) shows the transverse mass spectra at 40\AGeVc, compared with the NA49 results for central Pb+Pb collisions \cite{Afanasiev:2002mx}.
The spectra are approximately exponential; a deviation from the exponential function at low and high \mt in heavier systems indicates collective radial flow.

The \mt spectra are integrated to derive rapidity spectra, shown for central Ar+Sc collisions at six beam momenta in Fig.~\ref{fig:pim_spectra} (\emph{right}).
The spectra are well described by a sum of two symmetrically displaced normal distributions of independent amplitudes.

Multiplicities of pions of all charges $\avg{\pi}$ are calculated by integrating \pim rapidity spectra and using phenomenological dependences between multiplicities of pions of various charges~\cite{Naskret:2016lut}.
They are shown in Fig.~\ref{fig:pim_multiplicity}, compared with results from other experiments~\cite{Ahle:1998jc,Blobel:1975ka,Golokhvastov:2001pt}.
The values were divided by the average number of wounded nucleons $\avg{W}$.
For higher SPS energies the slope of the energy dependence is larger for the heavy systems (Pb+Pb, Ar+Sc) than for the light ones (p+p, Be+Be).
The Statistical Model of the Early Stage (SMES) predicts an increase of the slope at the onset of deconfinement due to the larger number of degrees of freedom in the quark-gluon plasma~\cite{Gazdzicki:1998vd}.

\subsubsection{Charged hadron spectra}
\begin{figure}
 \centering
  \includegraphics[height=5.2cm,trim={0.5cm 0 0 0},clip]{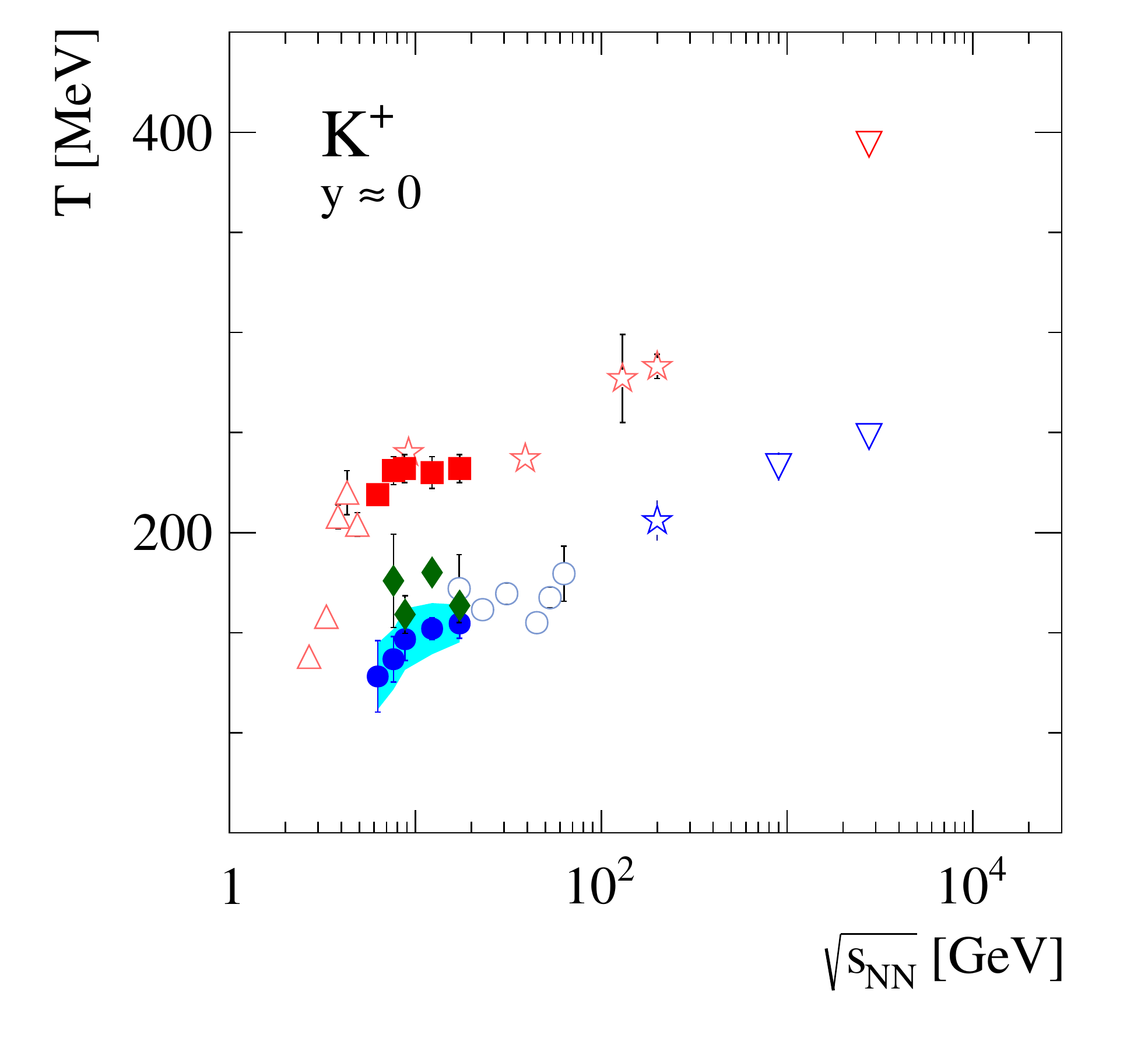}
  \includegraphics[height=5.2cm,trim={0.5cm 0 0 0},clip]{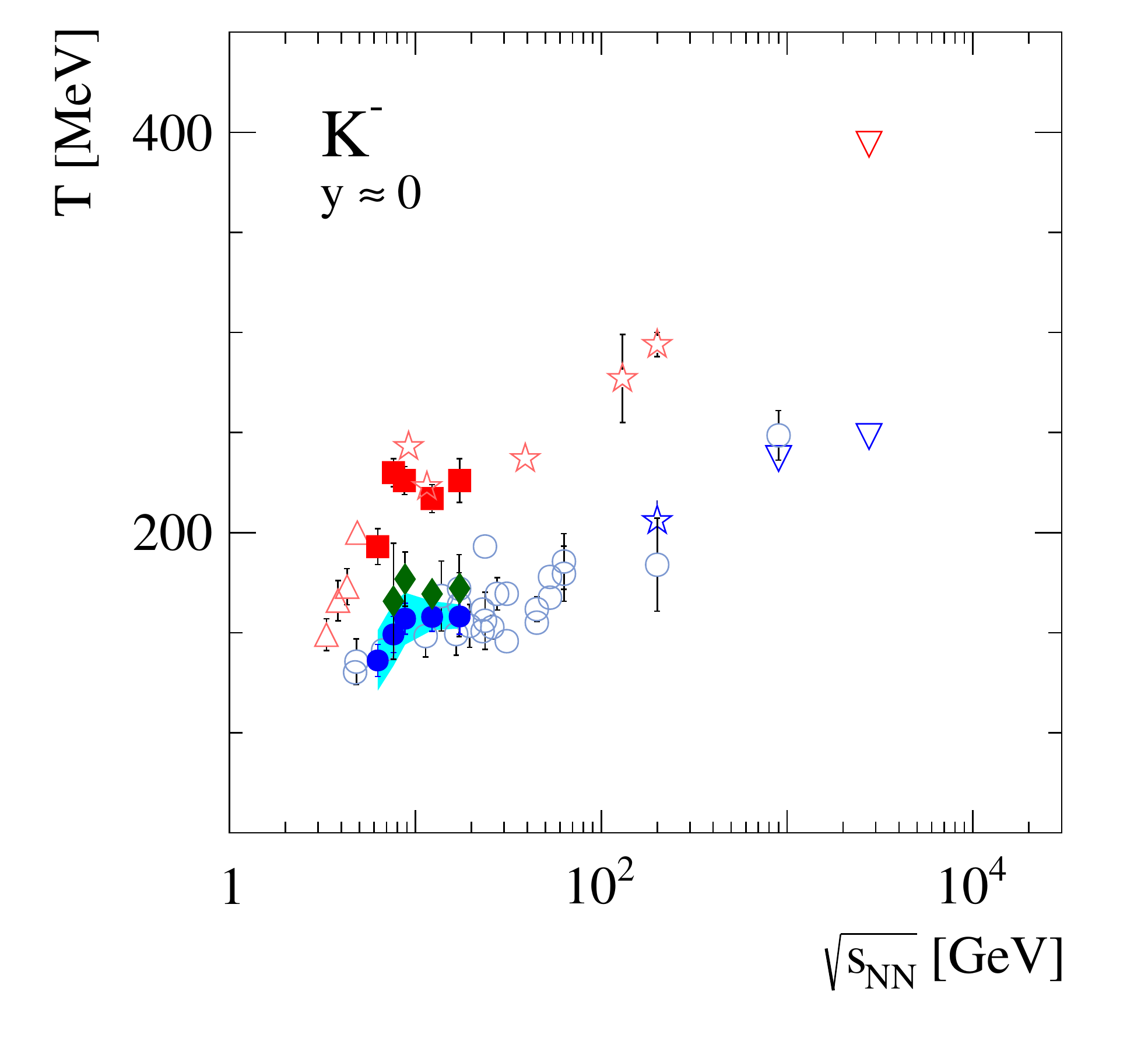}
  \includegraphics[height=5.2cm,trim={1.7cm 0 8.1cm 0},clip]{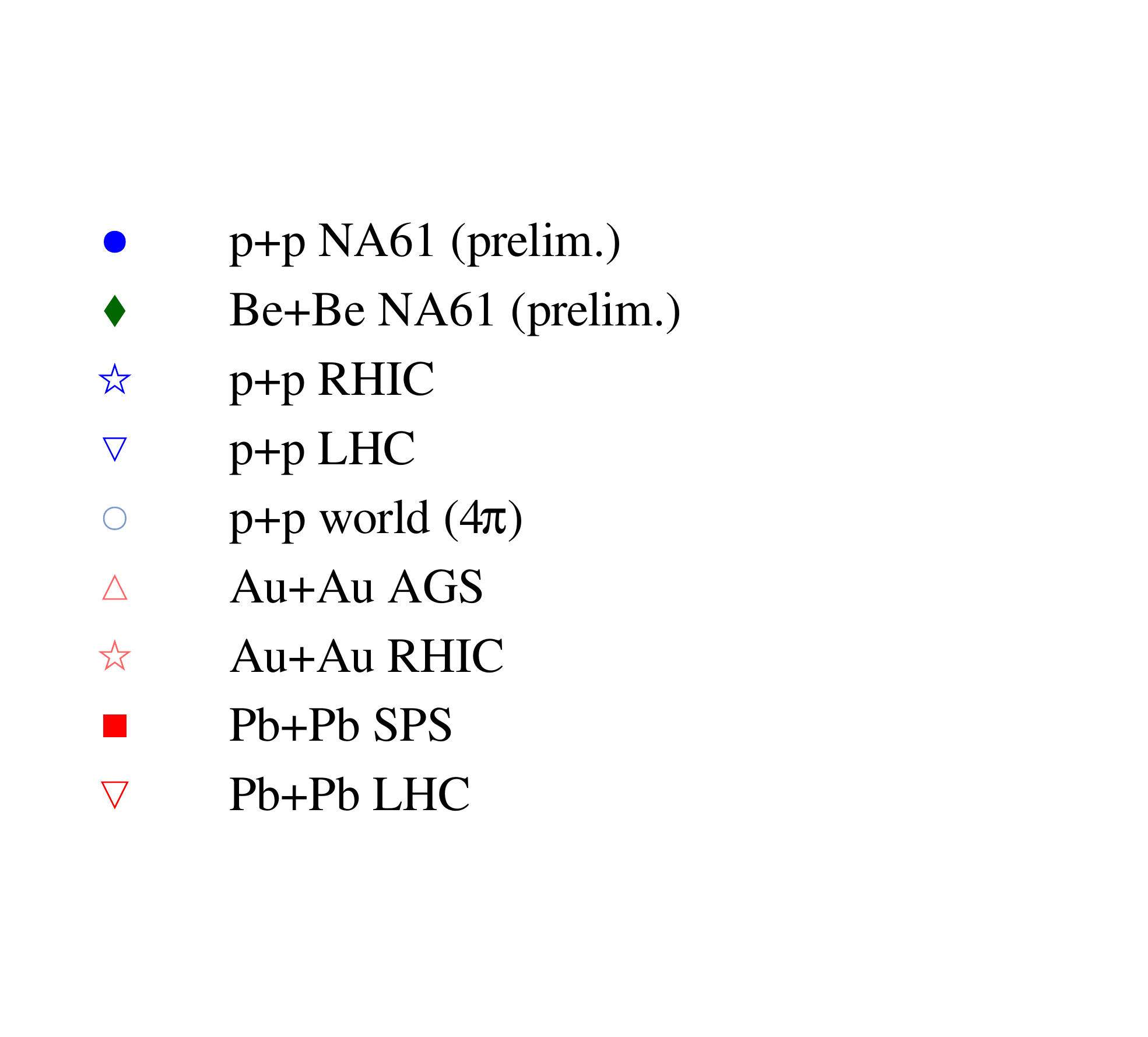}
 \caption{Energy dependence of the inverse slope parameter of the transverse mass distribution at mid-rapidity for charged kaons.
 The \NASixtyOne results on p+p interactions (full blue circles) and new results on Be+Be (full green diamonds) collisions are compared with world data on p+p and heavy ion (Pb+Pb and Au+Au) reactions.
 }\label{fig:step}
\end{figure}

\begin{figure}
 \centering
  \includegraphics[height=5.2cm,trim={0.5cm 0 0 0},clip]{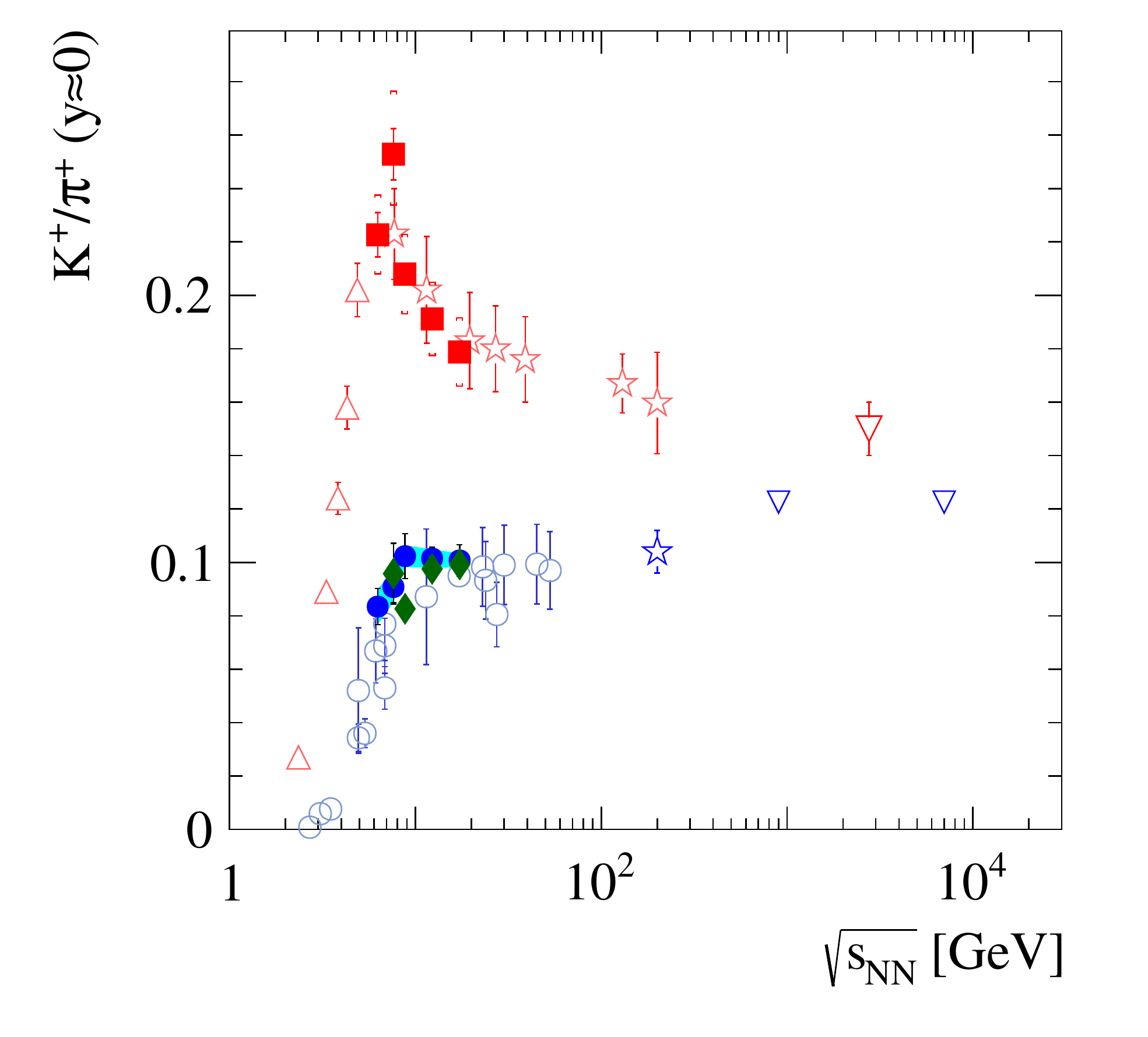}
  \includegraphics[height=5.2cm,trim={0.5cm 0 0 0},clip]{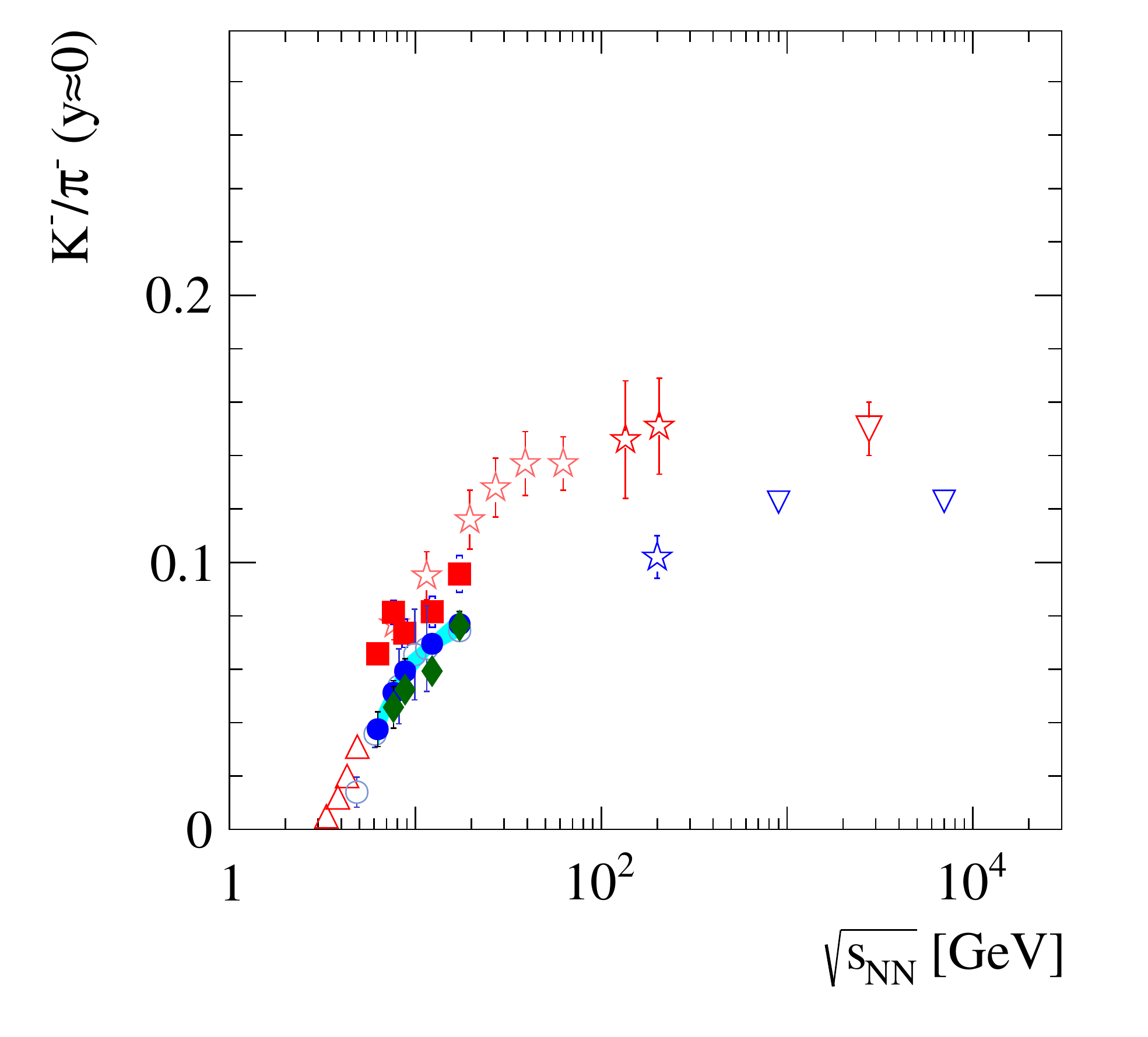}
  \includegraphics[height=5.2cm,trim={1.7cm 0 8.1cm 0},clip]{pic/legenda}
  \caption{Energy dependence of the positively (\emph{left}) and negatively (\emph{right}) charged kaon multiplicity divided by corresponding charged pion multiplicity at mid-rapidity.}\label{fig:horn}
\end{figure}

$\pi^\pm$, p, \pbar (in p+p interactions) and  K$^\pm$ (in p+p and central Be+Be collisions) were identified based on measurements of the energy loss in the TPCs (\dedx) and time of flight in the ToF detectors.
Figures~\ref{fig:step} and~\ref{fig:horn} present the energy dependence of the inverse slope parameter of the transverse mass distribution of charged kaons and the ratio of charged kaon to pion multiplicity at mid-rapidity, respectively.
The \NASixtyOne results for p+p~\cite{Pulawski:2015tka} and Be+Be interactions are compared to those from central Pb+Pb collisions of NA49~\cite{Alt:2007aa, Afanasiev:2002mx} and other experiments~\cite{Gazdzicki:1995zs,Gazdzicki:1996pk,Kliemant:2003sa,Arsene:2005mr,Aamodt:2011zj,Abelev:2008ab,Abelev:2014laa,Abelev:2012wca}.
For Pb+Pb collisions in the SPS energy range the local plateau ("step") in the inverse slope parameter visible in Fig.~\ref{fig:step} and the peak ("horn") in the \emph{left} panel of Fig.~\ref{fig:horn} were predicted by the SMES as signatures of the onset of deconfinement.

The \NASixtyOne results on p+p and Be+Be interactions greatly improve the quality of the available data on small systems.
They also reveal rapid changes of the energy dependence in the SPS energy range suggesting that some properties of hadron production previously attributed to onset of deconfinement in heavy ion collisions are present also in p+p interactions.
Interestingly while the inverse slope parameter in Be+Be collisions lies slightly above the one in p+p interactions, the values of the charged kaon to pion ratio are very close in Be+Be and p+p.

\subsubsection{Lambda spectra}
\begin{figure}
 \centering
\includegraphics[width=0.40\textwidth]{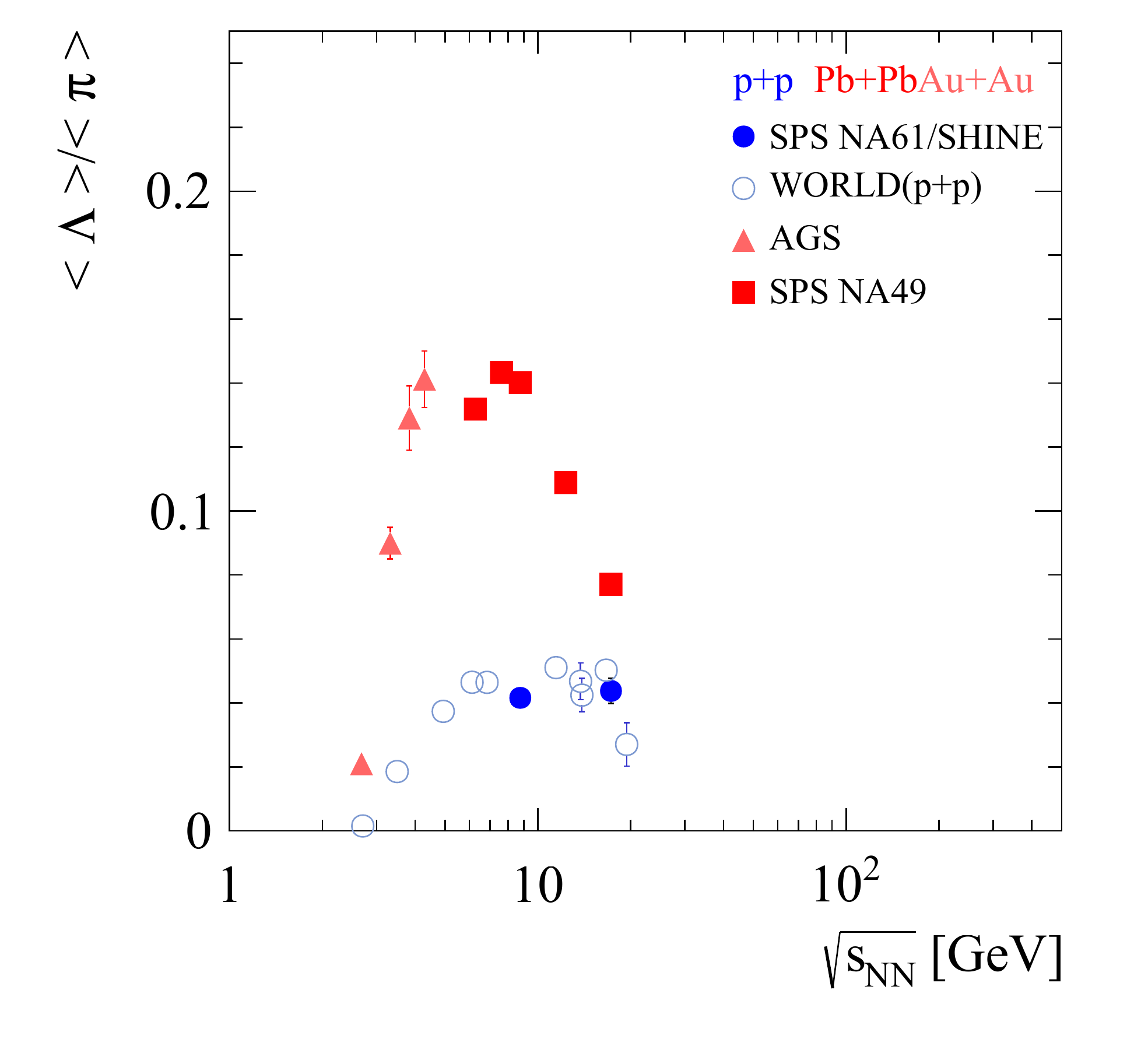}
\includegraphics[width=0.40\textwidth]{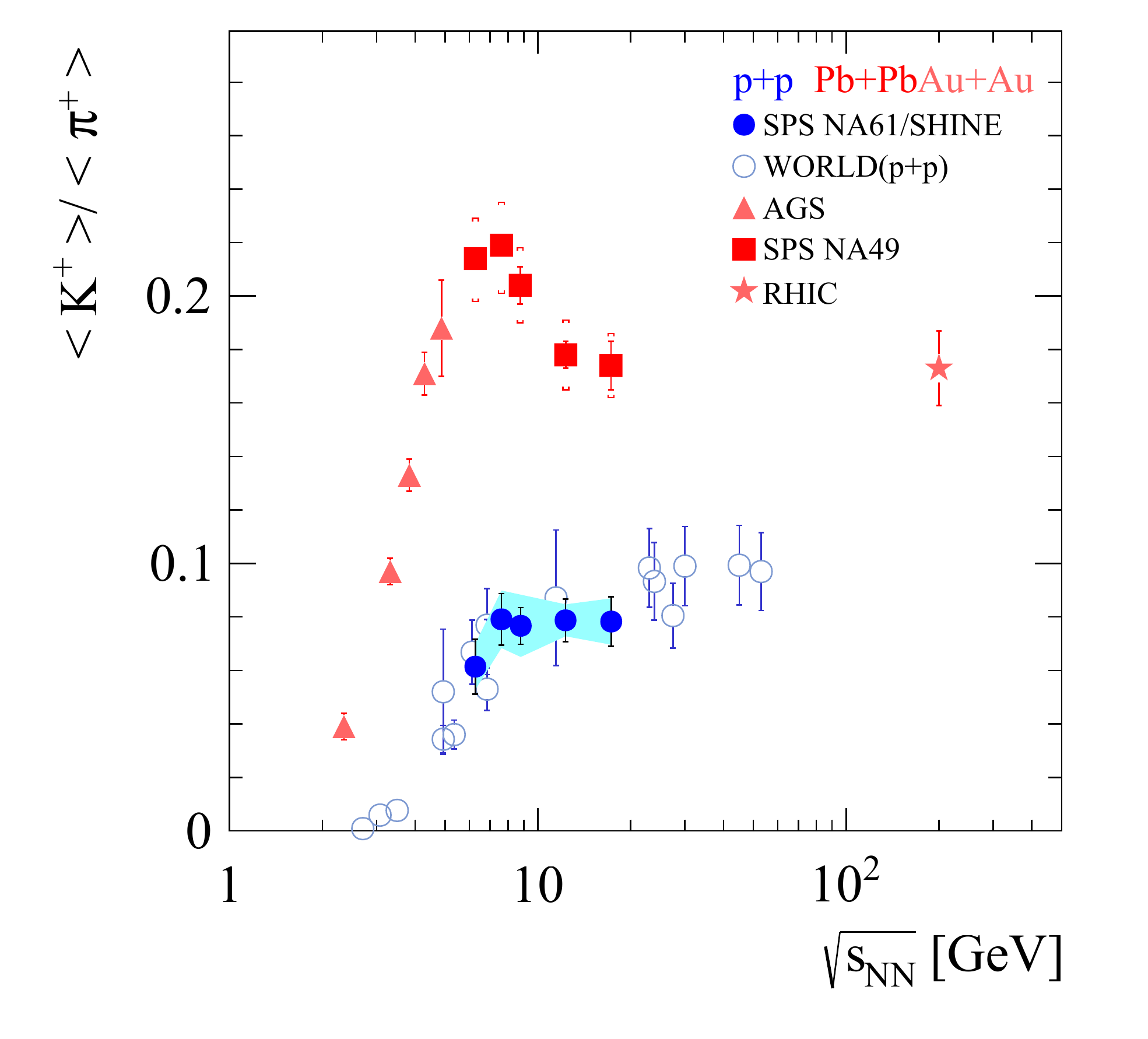}
\caption{
 \emph{Left:} \NASixtyOne measurement of the energy dependence of the ratio of total $\Lambda$ to $\pi$  multiplicity in p+p collisions at 40 and 150\GeVc (full blue circles) compared with world results on p+p and Pb+Pb collisions.
 \emph{Right:} \NASixtyOne measurement of the energy dependence the multiplicity ratio \kp/\pip in p+p collisions at 20--158\GeVc (full blue circles) compared with world results on p+p and Pb+Pb collisions.
}\label{fig:lambda}
\end{figure}

\NASixtyOne measured $\Lambda$ spectra in p+p interactions at 40~\cite{Herbert_SQM2016} and 158\GeVc~\cite{Aduszkiewicz:2015dmr}.
Figure~\ref{fig:lambda} (\emph{left}) presents the energy dependence of the ratio of total $\Lambda$ to $\pi$ multiplicity, compared with other results for p+p and heavy ion collisions.
For Pb+Pb collisions this energy dependence shows a maximum in the SPS energy range for ion collisions but not for p+p reactions.
A similar maximum is visible in the ratio of total \kp to \pip multiplicities, shown Fig.~\ref{fig:lambda}~(\emph{right}).
The observations for Pb+Pb collisions are consistent with the SMES prediction on the energy dependence of the strangeness to entropy ratio at the onset of deconfinement.

\subsection{Search for the critical point}
\subsubsection{Event-by-event fluctuations}
\NASixtyOne searches for the critical point by searching for non-monotonic dependences in event-by-event fluctuations of hadron production properties.
Results on two fluctuation measures will be presented:
\begin{itemize}
  \item Scaled variance of the multiplicity distribution $\omega[N] \equiv Var(N)/\avg{N}$, an intensive variable, insensitive to the system volume (size), but sensitive to volume fluctuations.
  \item $\Sigma[P_\textup{T}, N]$ measure of fluctuations of the transverse momentum and multiplicity, a strongly intensive variable, insensitive both to the system volume and its fluctuations.
\end{itemize}

Comparison of results of fluctuation measurements between various experiments is challenging due to differences in acceptance, volume fluctuations and choice of measures of fluctuations.
For this reason only \NASixtyOne results will be presented in this section.

\subsubsection{Multiplicity fluctuations}
\begin{figure}
 \centering
 \raisebox{2mm}{\includegraphics[width=0.49\textwidth]{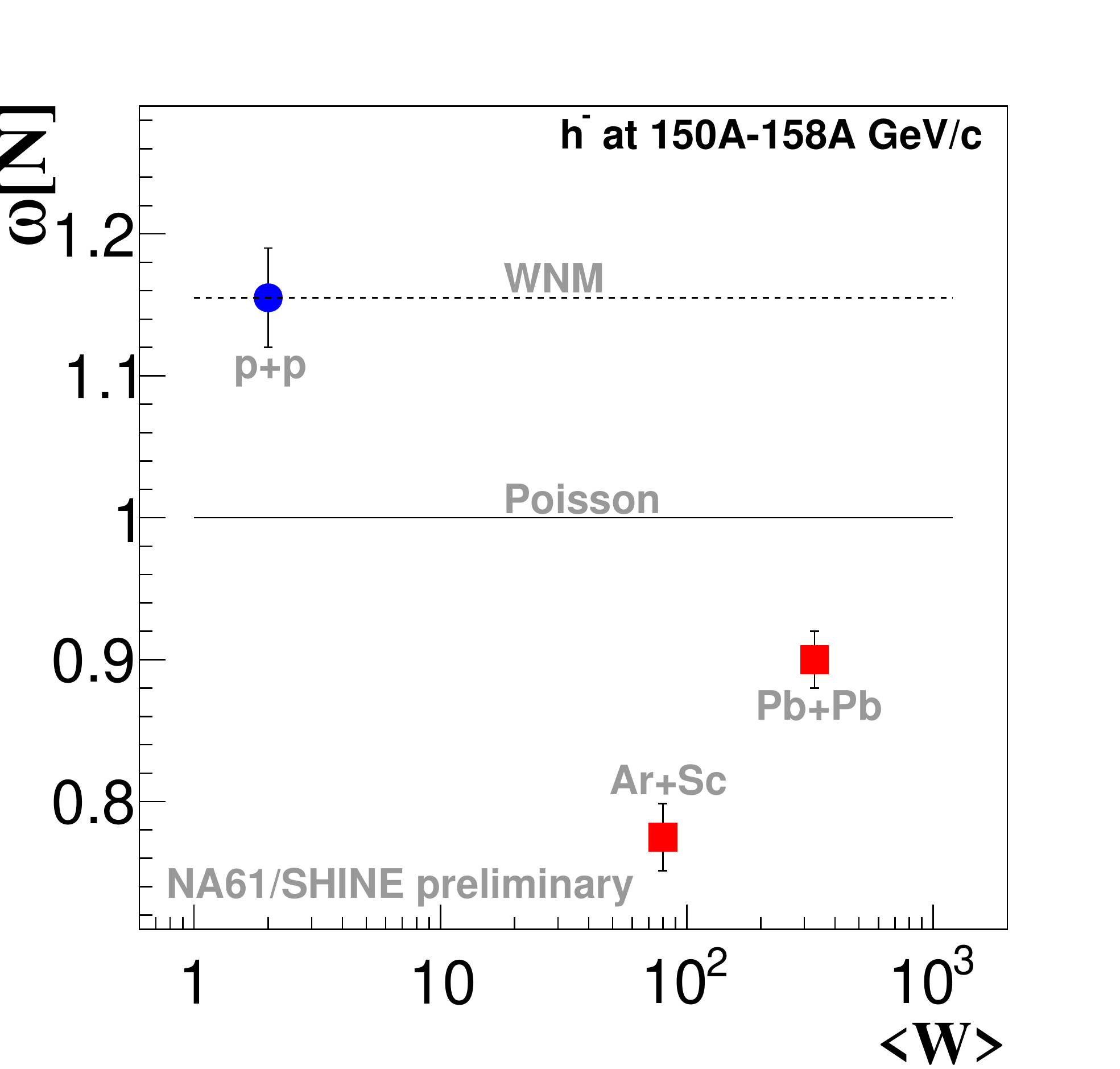}}
\includegraphics[width=0.49\textwidth]{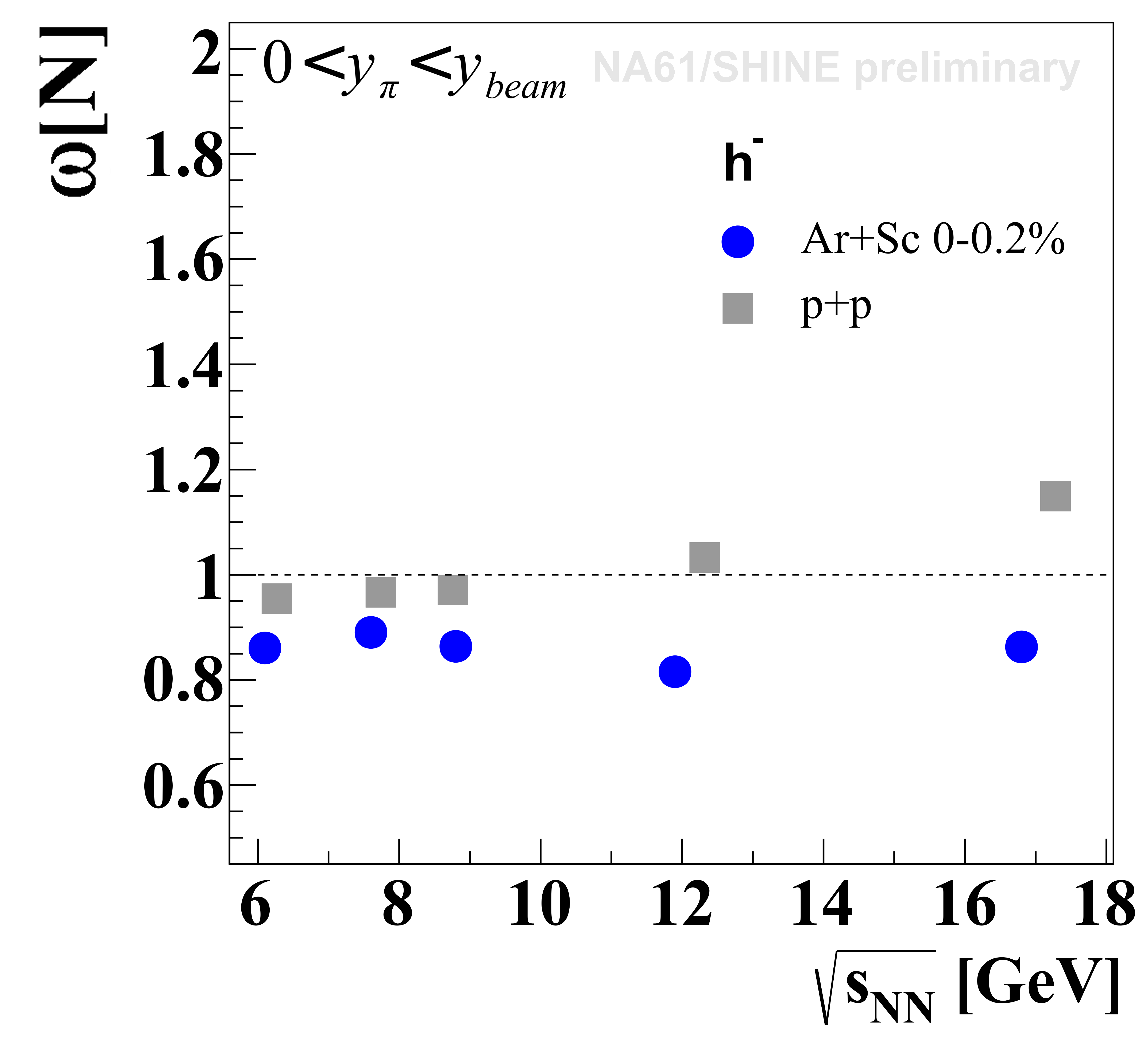}
 \caption{Scaled variance $\omega[N]$ of negatively charged hadron multiplicity distributions for p+p and 0.2\% most central Ar+Sc collisions measured by \NASixtyOne and 1\% most central Pb+Pb collisions measured by NA49.
   \emph{Left:} System size dependence of results calculated in the NA49 acceptance.
   The Wounded Nucleon Model prediction of the minimal $\omega[N]$ value in A+A collisions is indicated by the dashed line.
   The value of $\omega[N] = 1$ for the Poisson distribution is also marked.
   \emph{Right:} Energy dependence of $\omega[N]$ measured in the \NASixtyOne acceptance.
 }\label{fig:omega}
\end{figure}

Figure~\ref{fig:omega} (\emph{left}) shows the scaled variance $\omega[N]$ of the multiplicity distributions for p+p, central Ar+Sc, and Pb+Pb collisions at 150/158\AGeVc calculated in the NA49 acceptance~\cite{CPOD2016_Andrey}.
Only the 0.2\% most central Ar+Sc collisions were used in the analysis in order to eliminate volume fluctuations.
Results contradict the Wounded Nucleon Model prediction that the values for heavy systems will be greater or equal to those for p+p.

Figure~\ref{fig:omega} (\emph{right}) shows the energy dependence of the scaled variance in p+p and central Ar+Sc collisions calculated in the \NASixtyOne acceptance.
The Ar+Sc points lie systematically below the p+p ones and no indication for non-monotonic behaviour is visible.

\subsubsection{Transverse momentum and multiplicity fluctuations}
\begin{figure}
 \centering
\includegraphics[width=0.49\textwidth]{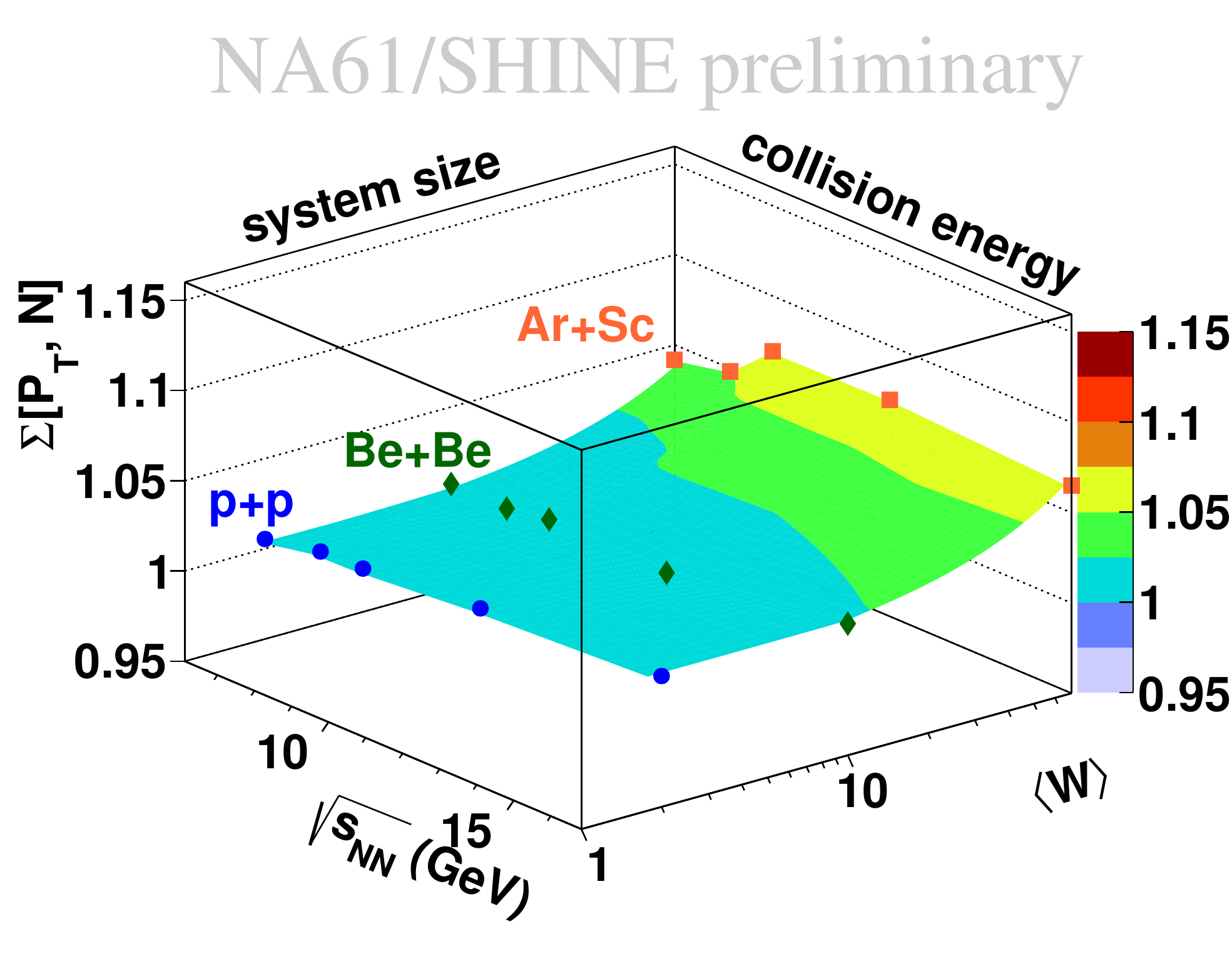}
\caption{Strongly intensive measure $\Sigma[P_\textup{T}, N]$ of the transverse momentum and multiplicity fluctuations of negatively charged hadrons measured at 5 beam momenta for p+p, Be+Be and Ar+Sc collisions.
 The estimated systematic uncertainty is of order of 5\%, similar to the magnitude of the visible variations.
 }\label{fig:sigma}
\end{figure}

Figure~\ref{fig:sigma} presents the energy and system size dependence of the $\Sigma[P_\textup{T}, N]$ measure of transverse momentum and multiplicity fluctuations~\cite{Grebieszkow:2016cza,CPOD2016_Evgeny}.
The results show no beam momentum dependence.
The 5\% increase from p+p to Ar+Sc is consistent with the estimated magnitude of the systematic uncertainty.
The Independent Particle Production Model predicts $\Sigma[P_\textup{T}, N] = 1$, which is consistent with the presented data~\cite{Gazdzicki:2013ana}.

\subsubsection{Higher order moments of the net-charge distribution in p+p collisions}
\begin{figure}
 \centering
\includegraphics[width=0.49\textwidth]{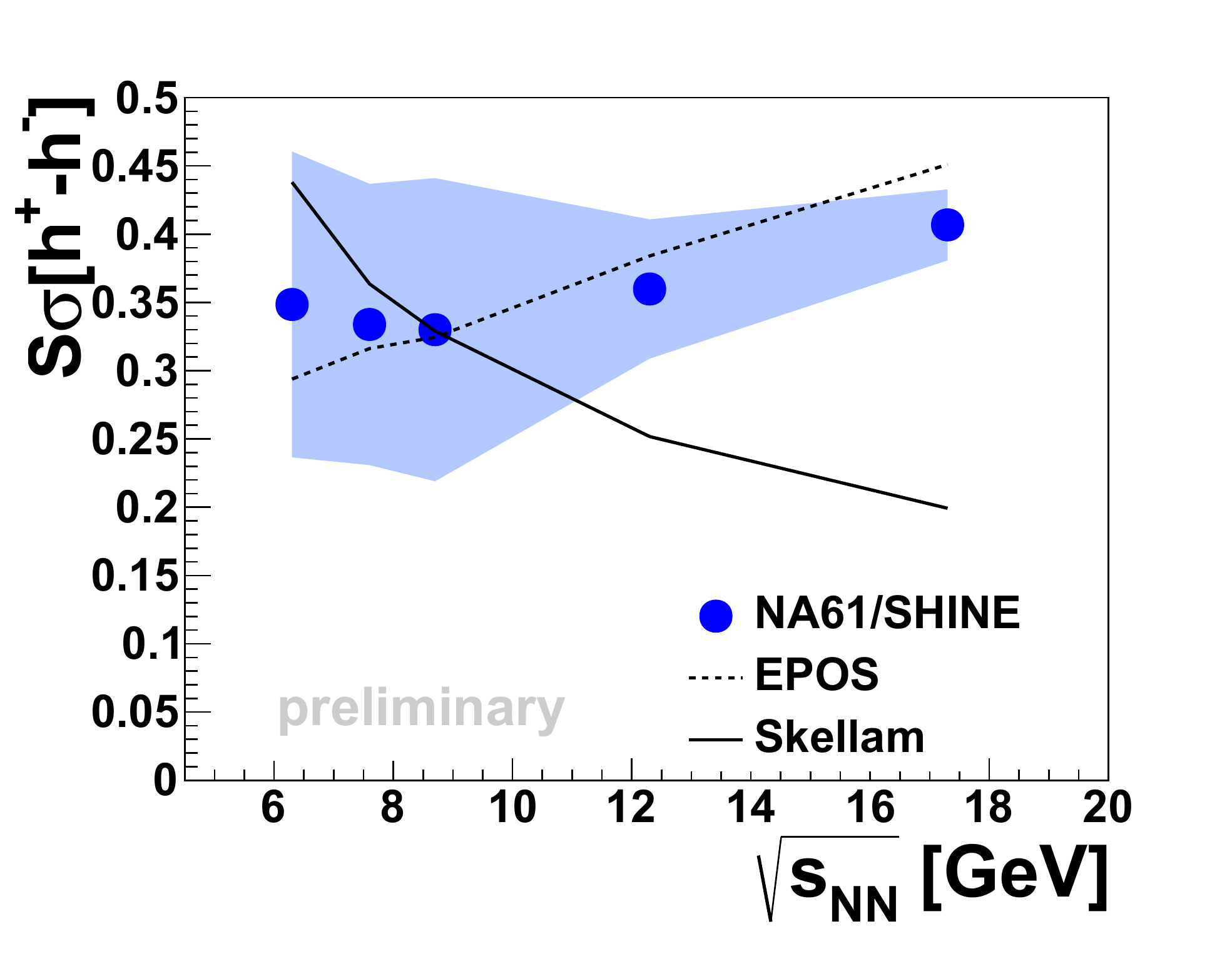}
\includegraphics[width=0.49\textwidth]{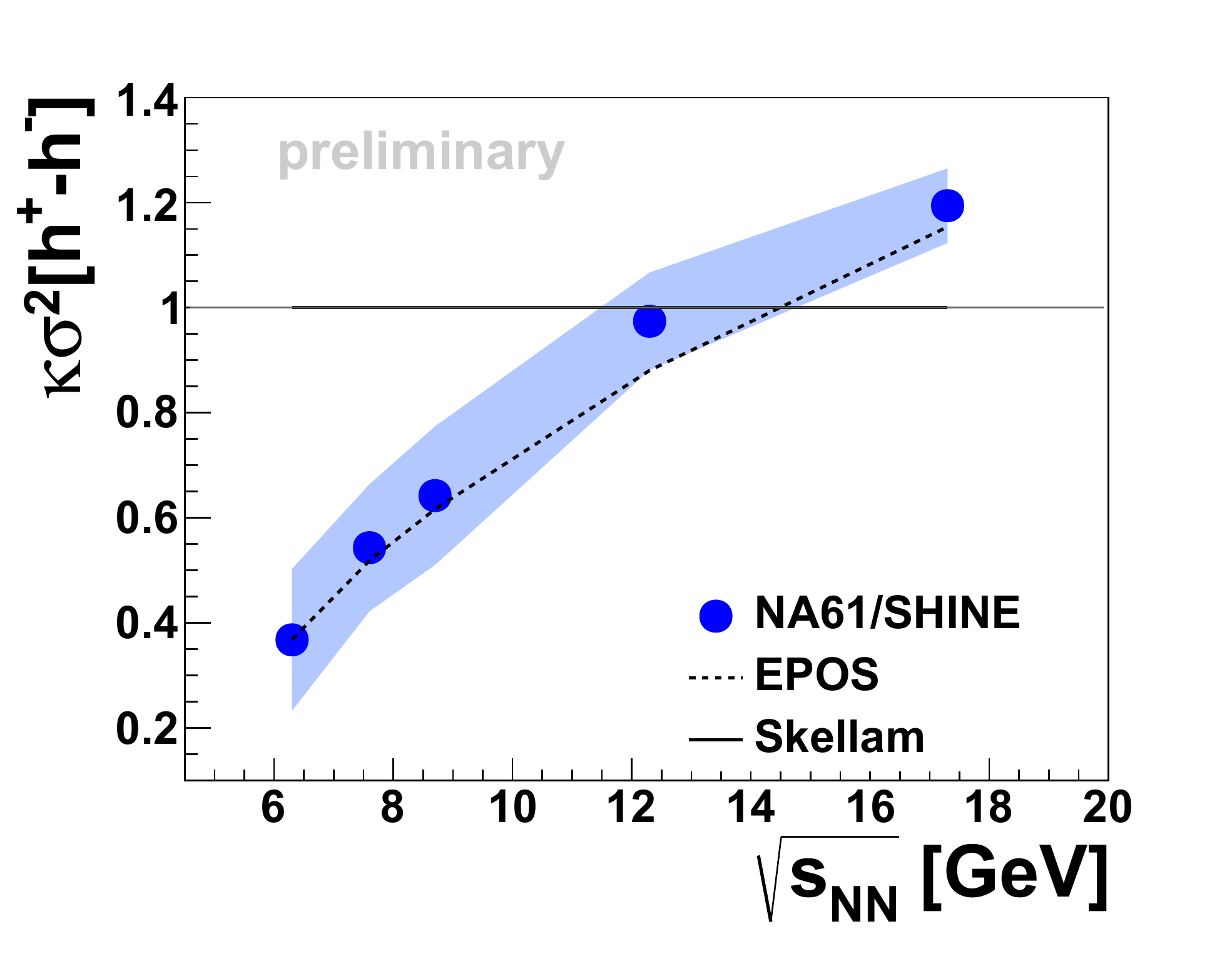}
\caption{Energy dependence of skewness $S$ multiplied by variance $\sigma$ (\emph{left}) and kurtosis $\kappa$ multiplied by $\sigma^2$ (\emph{right}) of the net-charge distribution measured in p+p interactions.}\label{fig:higher_orders}
\end{figure}

Higher order moments of multiplicity distributions might be particularly sensitive to effects of a critical point.
Figure~\ref{fig:higher_orders} shows the energy dependence of skewness and kurtosis of the net-charge distribution in p+p interactions~\cite{CPOD2016_Maja}.
No non-monotonic dependence is observed, but these results establish a reference for future measurements in collisions of heavier systems.

\section{\NASixtyOne now and in the future}
\subsection{\NASixtyOne in 2017--2018}
The \NASixtyOne two-dimensional system size and beam momentum scan will be completed with measurements of p+Pb, Xe+La and Pb+Pb collisions in 2017 and 2018.
\NASixtyOne with the new small-acceptance Vertex Detector will perform pilot open charm production measurements.
Moreover, precise measurements of fluctuations and collective effects in Pb+Pb collisions~\cite{PbAddendum} will be carried out.

\subsection{\NASixtyOne in 2021--2024}
A detector upgrade of \NASixtyOne is planned during the Long Shutdown 2 at CERN in years 2019--2020:
the readout speed will be increased to 1 kHz and a Large Acceptance Vertex Detector will be constructed.

The upgraded detector will allow the performance of a high statistics beam momentum scan with Pb+Pb collisions for precise measurements of open charm and multi-strange hyperon production in 2021--2024.
This program will complement future measurements at NICA, FAIR and J-PARC

\NASixtyOne also conducts extensive and precise particle production measurements for the neutrino physics program which are planned to be continued after 2020.

\section{Summary}
This contribution discusses recent results from the \NASixtyOne energy and system size scan performed to study the onset of deconfinement and to search for the critical point.
Results on particle spectra and fluctuations were presented.
New charged kaon spectra in Be+Be collisions at 30$A$--150\AGeVc were shown.
Surprisingly, also p+p reactions show rapid changes in particle production properties, partly resembling those seen in Pb+Pb collisions.
Results from Be+Be collisions are close to those from p+p reactions.
The charged kaon to pion ratios are consistent and the inverse slope parameters of the transverse mass distributions are marginally higher. 

Present results on fluctuations show no indication for non-monotonic dependence and thus no indication for a critical point.
Still such features may be revealed in the future results on Xe+La and Pb+Pb collisions. 

The presently ongoing two-dimensional scan will be completed in 2018.
As an extension of this program \NASixtyOne plans to measure precisely open charm and multi-strange hyperon production in 2021--2024.

\section{Acknowledgments:}
This  work  was  partially  supported  by  the  National  Science Centre, Poland grant 2015/18/M/ST2/00125.

\section{References}

\bibliographystyle{elsarticle-num}
\bibliography{include/na61References}







\end{document}